\documentclass[epl,showpacs,twocolumn,showemail]{revtex4}
\usepackage{bbm}
\usepackage{mathrsfs}
\usepackage{epsfig,psfrag}
\usepackage{amsmath,amsfonts,amssymb}
\usepackage[usenames]{color}
\usepackage{color}
\usepackage{soul}
\usepackage[dvips, bookmarks, colorlinks=true, plainpages = false, citecolor = blue, linkcolor = blue, urlcolor = blue, filecolor = blue]{hyperref}
\def\be{\begin{equation}}
\def\ee{\end{equation}}
\def\bea{\begin{eqnarray}}
\def\eea{\end{eqnarray}}

\begin{document}

\preprint{draft}

\title{Three Dimensional Ising Model, Percolation Theory and Conformal Invariance}
\author{ Abbas Ali Saberi$^1$ }
\email{a$_$saberi@ipm.ir; ab.saberi@gmail.com }
\author{ Horr Dashti-Naserabadi$^2$}
\address {$^1$School of Physics, Institute for Research in Fundamental
Sciences (IPM), P.O.Box 19395-5531, Tehran, Iran \\$^2$ Department
of Physics, Sharif University of Technology, P.O. Box 11155-9161,
Tehran, Iran}
\date{\today}

\begin{abstract}
The fractal structure and scaling properties of a 2d slice of the 3d
Ising model is studied using Monte Carlo techniques. The percolation
transition of geometric spin (GS) clusters is found to occur at the
Curie point, reflecting the critical behavior of the 3d model. The
fractal dimension and the winding angle statistics of the perimeter
and external perimeter of the geometric spin clusters at the
critical point suggest that, if conformally invariant in the scaling
limit, they can be described by the theory of Schramm-L\"{o}wner
evolution (SLE$_{\kappa}$) with diffusivity of $\kappa=5$ and
$16/5$, respectively, putting them in the same universality class as
the interfaces in 2d tricritical Ising model. It is also found that
the Fortuin-Kasteleyn (FK) clusters associated with the cross
sections undergo a nontrivial percolation transition, in the same
universality class as the ordinary 2d critical percolation.
\end{abstract}


\maketitle


Conformal invariance plays the key role in the analytical
description of two-dimensional (2d) systems at their critical points
\cite{NC}. One of the approaches to study the critical behavior of
spin models such as \emph{q}-state Potts model ($q\leq4$) is the
geometrical approach based on percolation theory \cite{sahimi, SA}
and also finite-size scaling (FSS) relations supplied by conformal
field theory (CFT) \cite{BH, Geometric}. Fortuin-Kasteleyn (FK) bond
representation of Potts model gives a close relation between
percolation and thermal phase transition in which critical
singularities can be represented in terms of FK clusters \cite{FK}.
An FK cluster can be constructed from a "geometric" cluster $-$
i.e., a set of nearest neighbor sites of like states, by assigning a
bond between each pair of sites with a certain temperature-dependent
probability. The FK clusters always percolate right at the critical
temperature $T_c$. The percolation temperature for both FK and
geometric clusters of a \emph{q}-state Potts model coincide only in
two dimensions.\\ Critical behavior of the interfaces of these
clusters in the scaling limit is conjectured, and in some cases
proven, to be described by the theory of Schramm-L\"{o}wner
evolution (SLE$_\kappa$) \cite{schramm}. The diffusivity $\kappa$
denotes the classification of conformally invariant interfaces whose
value for the hull and the external perimeter (EP) of an FK cluster,
$\kappa$ and $\kappa'$ respectively, satisfy a duality relation
$\kappa\kappa'=16$ \cite{duplantier1} (for a review on SLE, see
\cite{SLE}). Another similar duality also holds in 2d, between the
hull of an FK cluster and the hull of a spin cluster. Considering
these two duality relations together, one finds that the hull of a
spin cluster and the EP of an FK cluster can be described with the
same $\kappa$. Many connections between SLE and various 2d systems
such as turbulence, spin glasses, growth models, sandpile models and
disordered Potts model have been recently noticed and discussed
\cite{applications}.

For the Ising model, equivalent to $q=2$ Potts model, in three
dimensions (3d) the percolation transition of geometric spin (GS)
clusters occurs at some temperature $T_p$ well below the Curie point
$T_c$ \cite{Muller-Krumbhaar}. Thus, contrary to the FK clusters,
the 3d geometric clusters do not capture the critical properties of
the model. The question that arises is how the FK and GS clusters in
a 2d cross sectional slice of the 3d Ising model behave and whether
they reflect the critical properties of the 3d model. This is the
main subject of the present letter.

In this letter we present the results of extensive Monte Carlo study
of 3d Ising model simulated by using Wolff's algorithm \cite{wolff}
based on single cluster update. We examine the FSS hypothesis for
some percolation observables on a 2d slice of the lattice,
indicating that the percolation threshold of the geometric clusters
coincides with the critical point. This is in agreement with the
observation of \cite{Dotsenko}, in which the authors have found a
power-law scaling behavior in the length distribution of the loops
formed by an arbitrary cross section of the boundaries of 3d GS
clusters at $T=T_c$. In this paper, we undertake a detailed and
systematic study of the statistical and percolation properties of
all GS clusters on an arbitrary cross section of the system. Note
that disjoint GS clusters on a 2d cross section may belong to a 3d
GS cluster. The scaling behavior of the GS clusters motivated us to
examine possible conformal invariance of the critical geometric
clusters which seem to undergo a continuous transition at the
threshold. We find that the perimeter and the EP of a GS cluster at
the critical point satisfy the duality relation and their fractal
dimensions and winding angle statistics are compatible, in the
scaling limit, with the family of conformally invariant curves i.e.,
SLE$_\kappa$ with $\kappa=5$ and $16/5$ respectively. This latter is
in the same universality class as interfaces in tricritical Ising
model in two dimensions.
\\ Our analysis of FK clusters on 2d slices also indicates that
these clusters undergo a nontrivial percolation transition at
$\beta^{FK}_c/ \beta_c=1.617(1)$, in the same universality class as
the classical bond percolation.

There exist some numerical investigations of 3d Ising model on
special geometries with specific boundary conditions (bc)
\cite{Janke, Blote} which confirm a linear relation between the
scaling dimension of the operators of 3d systems and their
correlation lengths. This relation has been given by Cardy
\cite{cardy} by exploiting the conformal group transformations.

\begin{figure}[t]\begin{center}
\includegraphics[scale=0.37]{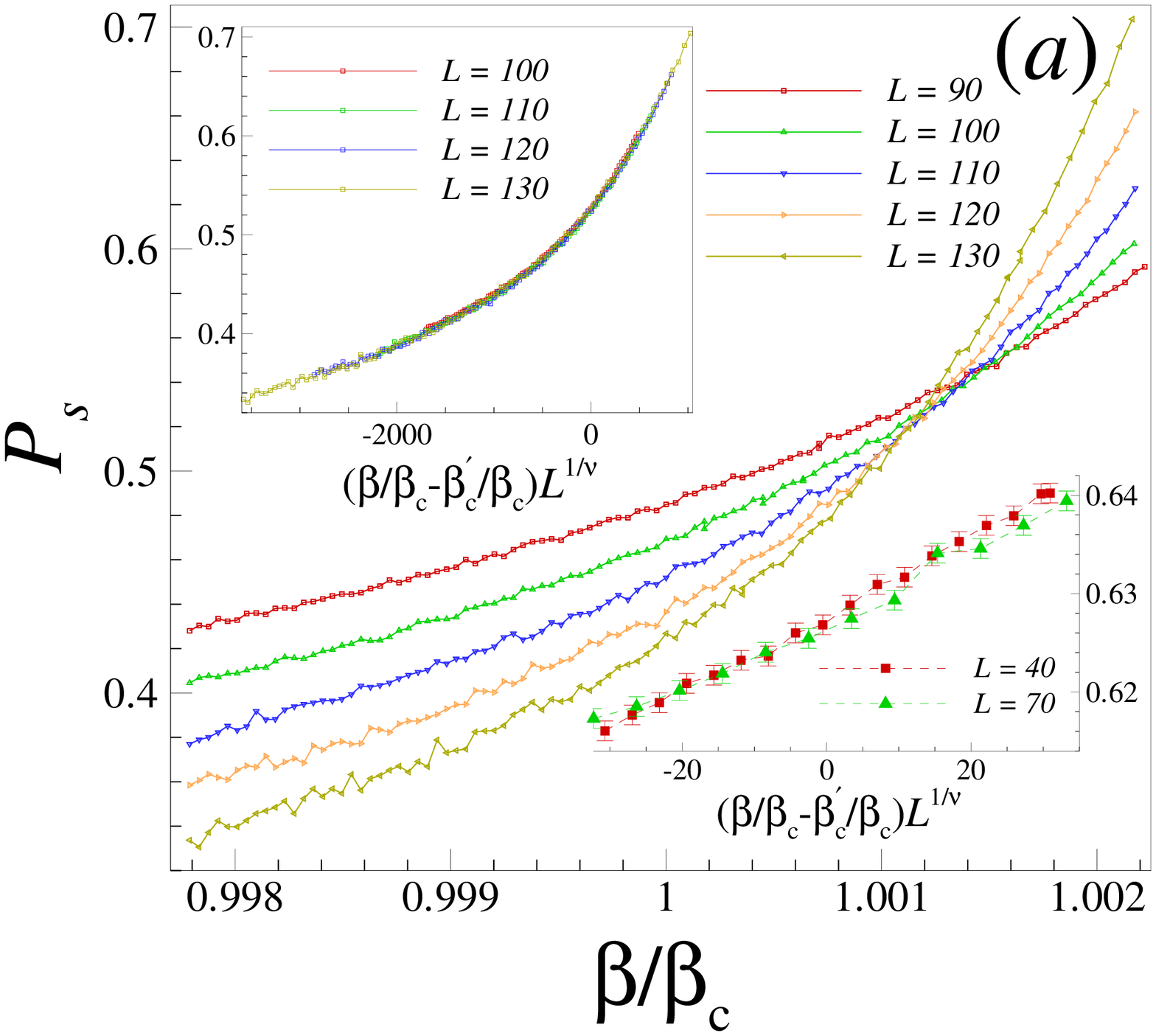}\\ \includegraphics[scale=0.37]{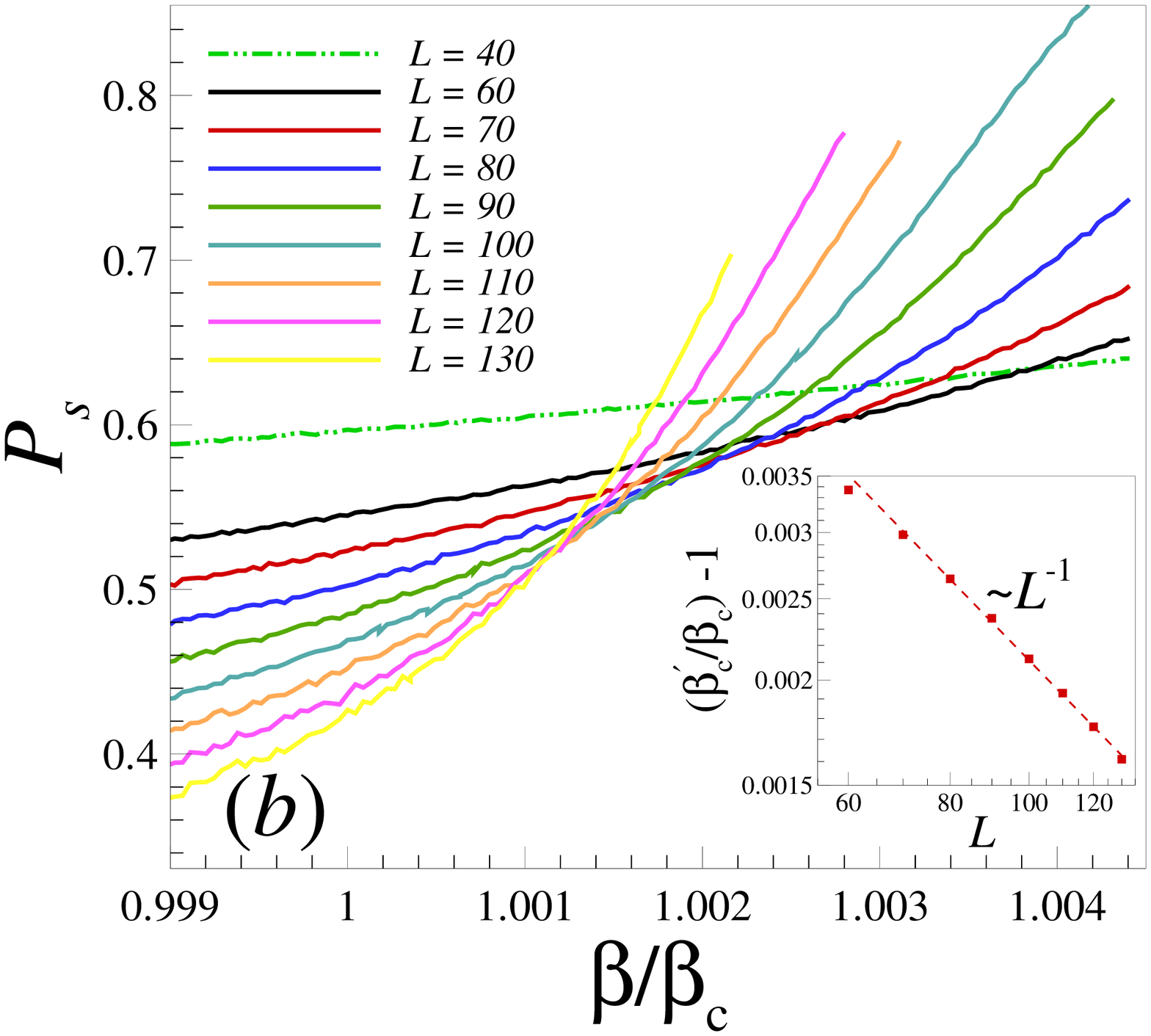}
\narrowtext\caption{\label{Fig1}(Color online) ($a$) The percolation
probability $P_s$ of GS clusters, measured on 2d cross sections of
3d Ising model of different size $L^3$, as a function of $\beta$.
Upper inset: FSS plots of the data with appropriate values
$\nu=0.36$ and $\beta'_c=1.001172 \beta_c$. Lower inset: the data
collapse very close to the crossing point for two $P_s$ curves of
sizes $L=40$ and $70$, with $\nu=0.36$ and $\beta'_c=1.003316
\beta_c$. ($b$) Size dependence of the crossing point
$\beta'_c(L,L')$ of different curves with the reference $P_s$ curve
of $L'=40$ (dotted-dashed line). }\end{center}
\end{figure}

\begin{figure}[b]\begin{center}
\includegraphics[scale=0.42]{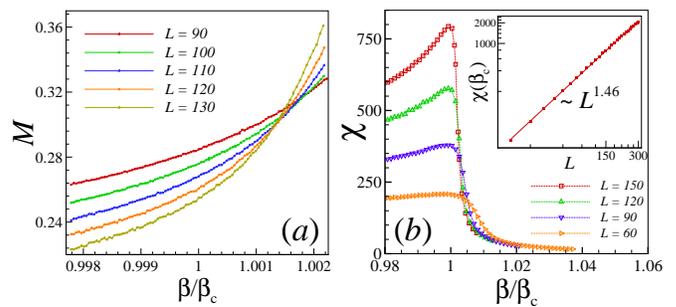}
\narrowtext\caption{\label{Fig2}(Color online) ($a$) The average
mass of the largest GS cluster $M$, and ($b$) the mean cluster size
$\chi$ as a function of $\beta$. Inset: size dependence of $\chi$ at
$\beta=\beta_c$.}\end{center}
\end{figure}

In order to investigate the fractal structure of the GS clusters on
2d cross sectional slices of 3d Ising model, we first examine the
FSS hypothesis for some percolation observables i.e., percolation
probability $P_s$, percolation strength $M$, and the mean cluster
size $\chi$. For this purpose, the simulation of 3d Ising model is
carried out by using the Wolff's Monte Carlo algorithm \cite{wolff}
on simple cubic (SC) lattices of different size $L^3$ with free bc,
and sizes up to $L= 150$. After discarding the first $L^3$ Monte
Carlo sweeps for equilibration, we then analyze the spin
configuration of a planar slice of the lattice parallel to the
\emph{xz}$-$plane located at $y=\lfloor L/2\rfloor$
($\lfloor\cdot\cdot\rfloor$ denotes the integer part). We check
whether it contains a GS cluster which spans the plane in just a
certain direction e.g., \emph{z}$-$direction. Therefore, we can
obtain the probability $P_s$ to have a spanning GS cluster at
temperature $T$. We then measure the average mass of the largest
spin-cluster, $M$. At each temperature, the averages are taken over
$10^2$ independent simulation runs and during each run another
average is taken over $10^5$ independent samples each of which is
taken after $L$ Monte Carlo steps \footnote{The dynamic critical
exponent $z=0.44(10)$ for the 3d Ising model simulated by the Wolff
algorithm has been reported in \cite{Tamayo}. To ensure that the
sample configurations are equilibrated and statistically
independent, we considered the relaxation time $\tau'\sim L^3$, and
the time scale between two selected samples $\tau\sim L$, much
larger than those reported in \cite{Tamayo}.}. This rather large
number of averages is needed to obtain smooth curves close enough to
the Curie point. The results are shown in Figs. \ref{Fig1} and
\ref{Fig2} as a function of inverse temperature $\beta$.\\As shown
in Fig. \ref{Fig1}, the curves $P_s(\beta)$ computed for different
cross sectional lattice sizes for sufficiently near sizes, cross at
the same point very close to the Curie point, implying that the
scaling dimension of the percolation probability is zero. This
agrees with what is expected from scaling theory \cite{BH} which
states that the $P_s$ curves should have the form $P_s(\beta) =
P_s[(\beta/\beta_c-1)L^{1/\nu}]$. The exponent $\nu$ characterizes
the divergence of the correlation length $\xi$ as the percolation
threshold is approached, $\xi\sim |\beta/\beta_c-1|^{-\nu}$. Note
that here $\xi$ is proportional to the spatial extent of the GS
clusters in 2d slices of the 3d Ising model \footnote{The two point
spin-spin correlation function in the original 3d Ising model is
equal to the pair connectedness function of 3d FK clusters, i.e.,
$\langle s(x)s(y)\rangle=\langle \delta_{C_x,C_y}\rangle$, which
equals the probability that points $x$ and $y$ belong to the same FK
cluster \cite{Hu}. The corresponding correlation length of the 3d
Ising model which is proportional to the spatial extent of the 3d FK
clusters is characterized by the well-known exponent $\nu\sim0.63$.
}.

By utilizing the data collapse, it is possible to estimate the
values of the exponent $\nu$ and the crossing point of the curves
$\beta'_c$. To measure the quality of the collapse of the curves we
define a function $S(\nu, \beta)$ as a function of the chosen values
of $\nu$ and $\beta$ (the smaller $S$ is indicative of a better
quality of the collapse $-$ see \cite{Kawashima, Bhattacharjee} and
appendix of \cite{HH}). We find its minimum $S_{min}\sim1.24$ for
values $\nu=0.36(3)$ and $\beta'_c=1.0012(10)\beta_c$, confirming
the percolation threshold of the 2d GS clusters at the Curie point.
Applying these values for $\nu$ and $\beta'_c$, rescaling of the
data for different lattice sizes fall onto a universal curve (see
upper inset of Fig. \ref{Fig1}($a$)). However crossing point of the
curves is slightly different from $\beta_c$, but if one does not
consider this discrepancy, the proper data collapse would not be
achieved because, due to the relatively small value of $\nu$, the
factor $L^{1/\nu}$ would magnify this discrepancy along the $x-$axis
and thus the curves do not collapse onto a single function. This
effect will be more evident when one zooms in around zero on the
$x-$axis. We find that the crossing point moves towards the Curie
point for larger system sizes (Fig. \ref{Fig1}($b$)). Considering
the $P_s$ curve for $L'=40$ as a reference curve, we find that the
crossing point of the other curves for different sizes with this
reference curve i.e., $\beta'_c(L,L')$, behaves like
$[(\beta'_c(L,40)/\beta_c)-1]\sim L^{-\omega}$, with
$\omega=1.02(5)$, which is independent of $L'$. Therefore, to obtain
a proper collapse even close enough to the crossing point, we need
to measure $\beta'_c$ with high precision which is possible for only
two $P_s$ curves. An example of such collapse is shown in the lower
inset of Fig. \ref{Fig1}($a$), for system sizes $L=40$ and $70$.
However, our estimation for the exponent $\nu$ is not accurate and
has to be reevaluated by considering the corrections-to-scaling
effects.

In order to check that the coincidence between the percolation
threshold $\beta'_c$ of the 2d GS clusters and the Curie point
$\beta_c$ is not specific to the considered lattice geometry, i.e.,
SC structure, we repeated the Monte Carlo computations for a
face-centered cubic (FCC) lattice. Note that both $\beta'_c$ and
$\beta_c$ depend on lattice geometry. We find again their
coincidence for the FCC lattice.

We have also computed the percolation strength $M$, as the order
parameter in the percolation, and the mean cluster size $\chi$ (Fig.
\ref{Fig2}). We find that, at the Curie point $\chi$ scales as
$L^{\gamma'}$, with $\gamma'=1.46(4)$. These are indicative of a
continuous transition of 2d GS clusters.\\These scaling arguments
suggest a fractal structure for 2d GS clusters, meaning that at
$\beta_c$ there exists a set of GS clusters and their surrounding
loops at all scales.

In order to investigate manifestation of conformal invariance in the
critical behavior of the GS clusters on the cross sections, we
analyze statistical properties of the boundaries of these clusters
and compare them with the SLE curves. We first measure the fractal
dimension of the cluster boundaries and check the duality relation.
The fractal dimension of SLE curves is given by $d_f=1+\kappa/8$,
and the duality conjecture \cite{duplantier1} states that the EP of
SLE hulls for $\kappa > 4$ looks locally as SLE curves but with a
dual value $\kappa'=16/\kappa$.

\begin{figure}[t]\begin{center}
\includegraphics[scale=0.39]{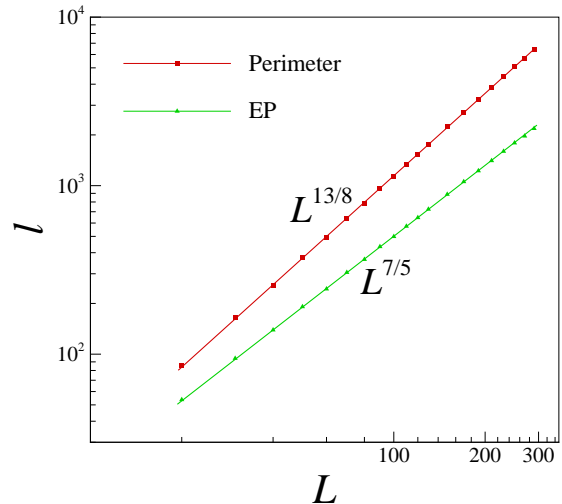}
\narrowtext\caption{\label{Fig3}(Color online) The average length of
the perimeter ($\blacksquare$) and EP ($\blacktriangle$) of a
spanning GS cluster versus the system size $L$, compared with ones
for SLE$_5$ and SLE$_{16/5}$ curves (solid lines), respectively. The
error bars are almost the same size as the symbols.}\end{center}
\end{figure}

We undertook extensive Monte Carlo simulations with anisotropic
geometry to obtain a reliable estimate of the fractal dimension at
the Curie point $\beta_c$ (the value of $\beta_c$ has been
determined to a high precision in Ref. \cite{T_c}). We present the
results of calculations for rather large system sizes of $L_x\times
L_y\times L_z$, with $L_y=L_z=L$, $L_x=4L$ and $10\leq L\leq 290$.
The periodic boundary condition was applied along all three
directions. After equilibration, we collected a number of $10^5$
($5\times 10^4$) of independent spin configurations on  a planar
cross section of the 3d sample located at $y=\lfloor L/2\rfloor$ for
smallest (largest) system size. Each configuration was taken after
the $L_x$ Monte Carlo steps. We then identified all GS clusters on
each cross section of a fixed size $L$, and marked all spanning
clusters (if exist) along the \emph{z}$-$direction. For each
spanning cluster, a walker was applied to determine each of its
perimeters which connect the lower boundary to the upper one along
the \emph{z}$-$direction. The walker was forced to start moving from
a node on the dual lattice (which is also a square lattice) lied at
the most left (right) site belonging to the spanning cluster on the
lower boundary, by using the turn-right (left) \emph{tie-breaking}
rule \cite{Saberi} and thereby, for each size $L$, an ensemble of
interfaces was obtained. It is shown in \cite{Saberi} that using
these tie-breaking rules, one can obtain the interfaces on a square
lattice without any ambiguity compatible with the conformal
invariant properties of critical lattice models. According to each
interface, the EP was also determined as the border of the
corresponding interface after closing off all its boundary narrow
passageways of lattice spacing $a$.

\begin{figure}[b]\begin{center}
\includegraphics[scale=0.39]{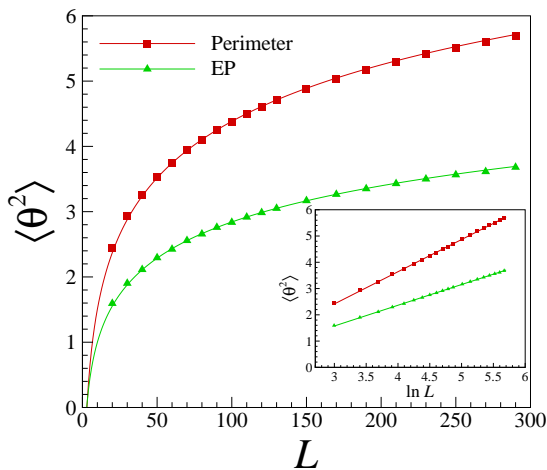}
\narrowtext\caption{\label{Fig4}(Color online) Variance of the
winding angle for the perimeter ($\blacksquare$) and EP
($\blacktriangle$) of GS clusters on a 2d cross section of the 3d
Ising model. The solid lines are set according to the Eq.
(\ref{winding}) for comparison, with $c\simeq-1.37$ and $\kappa=5$
for upper graph and with $c\simeq-0.84$ and $\kappa=16/5$ for the
lower graph. In the inset, the variance in semilogarithmic
coordinates.}\end{center}
\end{figure}

For fractal interfaces, their average length $l$, measured in units
of lattice spacing $a$ (which is set to unity here), scales as a
function of the system size $L$ as $l\sim a(L/a)^{d_f}$ where $d_f$
is the fractal dimension. Applying this relation for the ensemble of
the perimeters and the EPs, we find their fractal dimensions equal
to $d_f=1.621(10)$ and $d'_f=1.389(10)$, respectively. These results
give rise to $\kappa=4.96(8)$ and $\kappa'=3.11(8)$ for the
perimeter and EP of a GS cluster, respectively, comparable with ones
for SLE curves with diffusivity $\kappa=5$ and $\kappa'=16/5$, which
satisfy the duality relation (we find $\kappa\kappa'=15.42(65)$,
comparable with the exact one $\kappa\kappa'=16$). To show this
consistency, we have compared the results of our simulations with
those predicted for SLE curves in Fig. \ref{Fig3}. The exponent
$\delta'=1.23(1)$ obtained in \cite{Dotsenko} for the hull of the GS
clusters is also in a good agreement with our result
$\delta=2/d_f=2/(1+\kappa/8)\sim1.23077$ with $\kappa=5$.

To verify further the consistency between the statistical behavior
of the interfaces of GS clusters in a 2d cross section of the 3d
Ising model and the SLE curves with $\kappa=5$ and $\kappa'=16/5$,
let us now examine their winding angle statistics which is predicted
by the theory of SLE for critical interfaces \cite{schramm}. We use
the definition given in \cite{wilson}. For each interface, an
arbitrary winding angle is attributed to the first edge (which is
taken to be zero here). The winding angle for the next edge is then
defined as the sum of the winding angle of the present edge and the
turning angle to the new edge measured in radians. It is shown that
\cite{wilson} the variance in the winding grows with the system size
like \be\label{winding} \langle\theta^2\rangle = c+
\frac{\kappa}{4}\ln L,\ee where $c$ is a constant whose value is
irrelevant. As shown in Fig. \ref{Fig4}, the winding angle
statistics of GS interfaces are in a good agreement with those of
the corresponding SLE curves. We find that $\kappa= 4.93(10)$ and
$\kappa'= 3.15(10)$.

\begin{figure}[t]\begin{center}
\includegraphics[scale=0.37]{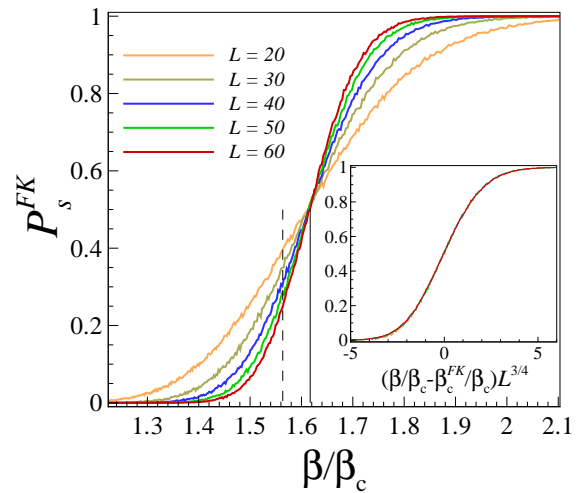}
\narrowtext\caption{\label{Fig5}(Color online) Percolation
probability for FK clusters, $P^{FK}_s$, on a 2d cross section of 3d
Ising model as a function of $\beta$, for different lattice size
$L$. Inset: the data collapse with $\nu=4/3$ for 2d classical
percolation, and $\beta^{FK}_c/ \beta_c=1.617$. }\end{center}
\end{figure}

The point which has to be explained and discussed here is the
seeming discrepancy between the exponents obtained from the direct
percolation properties of the GS clusters and the ones obtained by
analyzing the fractal properties of their hull and external
perimeter. The latter study gives rise to a connection between
statistical behavior of the boundaries of GS clusters in a 2d cross
section of the 3d Ising model with those of FK clusters in a pure 2d
tricritical Ising model, while the values of the geometrical
exponents obtained in the first part of the paper are obviously not
those predicted for tricritical Ising model at its thermal critical
point. Although we do not have a full description of such
discrepancy, its resolution seems to lie on natural difference
between GS and FK clusters. Moreover, the notion of conformal
invariance in the statistical behavior of the hull and external
perimeter of GS clusters requires extraction of the statistics of
the Loewner driving function and showing that it is a Brownian
motion, which is very hard to do on small samples such as ours.
These will be investigated in details in a future work.

So far, we have focused only on the statistical properties of GS
clusters. We now present the results of the same study for the FK
clusters built up on a 2d spin configuration by assigning a bond
between each pair of nearest neighbor sites of the like spins with
probability $p=1-e^{-2\beta}$. Our analysis indicates that the
percolation transition of the FK clusters occurs at $\beta^{FK}_c/
\beta_c=1.617(1)$ (shown by the vertical solid line in Fig.
\ref{Fig5}). This is obtained by computing the percolation
probability shown in Fig. \ref{Fig5}, with the same procedure as for
GS clusters. We find that the critical behavior of these clusters is
in the same universality class as the 2d critical percolation. As
shown in the inset of Fig. \ref{Fig5}, all curves fall well onto a
universal curve with the same correlation exponent i.e., $\nu=4/3$
as for classical bond percolation. The percolation threshold of 2d
bond percolation on a square lattice is located at $p=1/2$ which is
determined by the vertical dashed line in the Fig. \ref{Fig5}, at
$\beta<\beta_c^{FK}$, indicative of a nontrivial percolation
transition at $\beta_c^{FK}$. The latter is expected due to the
condensation of the majority spins at low temperatures. The analysis
of the statistical properties of the perimeter and EP of FK clusters
will appear in the forthcoming paper.

We would like to thank J. Cardy and M. Sahimi for their constructive
comments and  M.M. Sheikh-Jabbari for critical reading of the
manuscript. A.A.S. acknowledges financial support by the National
Elite Foundation (Bonyad-e-Melli-e-Nokhbegan) of Iran, and INSF
grant No. 87041917.

\end{document}